\documentclass[journal=jacsat,manuscript=article]{achemso}

\usepackage{subfig, graphicx}
\usepackage{verbatim}
\usepackage[version=3]{mhchem}
\usepackage{latexsym}
\usepackage{amsmath}
\usepackage{setspace}
\usepackage[nolist]{acronym}
\usepackage{multirow}
\usepackage{upgreek}
\usepackage{amssymb}
\usepackage{tipa}
\usepackage{dcolumn}
\usepackage{bm}
\usepackage{newunicodechar} 
\usepackage{tabularx}
\usepackage{makecell}
\usepackage{siunitx}
\usepackage{xcolor}
\usepackage{soul}

\DeclareSIUnit\mob{\cm\squared\per\volt\per\second}
\DeclareSIUnit\conc{\per\cubic\cm}
\DeclareSIUnit\angstrom{\text {Å}}

\author{Xinwei Wang}
\affiliation{ 
Department of Materials, Imperial College London, Exhibition Road, London SW7 2AZ, UK
}
\author{Alex M. Ganose}
\affiliation{ 
Department of Materials, Imperial College London, Exhibition Road, London SW7 2AZ, UK
}
\author{Seán R. Kavanagh}
\affiliation{ 
Department of Materials, Imperial College London, Exhibition Road, London SW7 2AZ, UK
}
\alsoaffiliation{ 
Thomas Young Centre and Department of Chemistry, University College London, 20 Gordon Street, London WC1H 0AJ, UK
}
\author{Aron Walsh}
 \email{a.walsh@imperial.ac.uk}
\affiliation{ 
Department of Materials, Imperial College London, Exhibition Road, London SW7 2AZ, UK
}

\title{Band Versus Polaron: Charge Transport in Antimony Chalcogenides}

\begin{document}

\begin{abstract}
Antimony sulfide (\ce{Sb2S3}) and selenide (\ce{Sb2Se3}) are emerging earth-abundant absorbers for photovoltaic applications. Solar cell performance depends strongly on charge carrier transport properties, but these remain poorly understood in \ce{Sb2X3} (X = S, Se). Here we report band-like transport in \ce{Sb2X3} by investigating the electron-lattice interaction and theoretical limits of carrier mobility using first-principles density functional theory and Boltzmann transport calculations. We demonstrate that transport in \ce{Sb2X3} is governed by large polarons with moderate Fröhlich coupling constants ($\alpha$ $\approx$ 2), large polaron radii (extending over several unit cells) and high carrier mobility (an isotropic average of \textgreater \SI{10}{\mob} for both electrons and holes). The room temperature mobility is intrinsically limited by scattering from polar phonon modes and is further reduced in highly defective samples. Our study confirms that the performance of \ce{Sb2X3} solar cells is not limited by intrinsic self-trapping.
%
\end{abstract}


\maketitle

\begin{acronym}
\acro{PV}{photovoltaic}
\acro{PCE}{power conversion efficiency}
\acro{PCEs}{power conversion efficiencies}
\acro{GBs}{grain boundaries}
\acro{DLTS}{deep-level transient spectroscopy}
\acro{ODLTS}{optical deep-level transient spectroscopy}
\acro{TAS}{thermal admittance spectroscopy}
\acro{ITS}{isothermal transient spectroscopy}
\acro{SQ}{Shockley–Queisser}
\acro{SRH}{Shockley–Read–Hall}
\acro{vdW}{van der Waals}
\acro{VASP}{Vienna Ab initio Simulation Package}
\acro{DFT}{density functional theory}
\acro{1D}{one-dimensional}
\acro{2D}{two-dimensional}
\acro{3D}{three-dimensional}
\acro{DOS}{density of states}
\acro{VB}{valence band}
\acro{CB}{conduction band}
\acro{VBM}{valence band maximum}
\acro{CBM}{conduction band maximum}
\acro{PESs}{potential energy surfaces}
\acro{DFPT}{density functional perturbation theory} 
\acro{ADP}{acoustic phonons}
\acro{IMP}{ionized impurities} 
\acro{POP}{polar optical phonons} 
\acro{LO}{longitudinal optical}
\acro{SIE}{self-interaction error}
\acro{PL}{photoluminescence}
\acro{PIA}{photoinduced absorption}
\acro{TA}{transient absorption}
\end{acronym}

~\
Antimony chalcogenides (\ce{Sb2X3}; X=S, Se) have emerged as promising light absorbing materials due to their attractive electronic and optical properties, including ideal band gaps (\SIrange{1.1}{1.8}{\electronvolt}) and high optical absorption coefficients (\textgreater \SI{e5}{\per\cm})\cite{versavel2007structural,liu2016green,messina2009antimony,lai2012preparation,chen2015optical,vadapoo2011self,vadapoo2011electronic,nasr2011electronic,savory2019complex,lei2019review,wang2022lone}. 
They are binary compounds with earth-abundant, low-cost and non-toxic constituents.
The \ac{PCEs} in \ce{Sb2X3} solar cells have improved rapidly over the past decade, with record efficiencies reaching \SI{7.50}{\percent} and \SI{10.12}{\percent} for \ce{Sb2S3} and \ce{Sb2Se3}, respectively\cite{choi2014highly,duan2022effi}. Nevertheless, efficiencies are still well below those seen in state-of-the-art CdTe or hybrid halide perovskite devices, which have reached above \SI{25}{\percent} under laboratory conditions\cite{green2021solar}.

The underlying efficiency bottleneck is unclear. While the structural, electronic and optical properties of \ce{Sb2X3} have been widely investigated, 
the charge carrier dynamics, which critically affect conversion efficiencies, remain controversial. Charge carrier transport in \ce{Sb2X3} has been reported by several studies\cite{yang2019ultrafast,grad2021charge,zhang2021suppressing,grad2020photoexcited,chen2017characterization}, but there are several fundamental questions that remain unanswered. The first is whether the nature of carrier transport is band-like or thermally-activated hopping. \citet{yang2019ultrafast} studied the charge carrier dynamics in \ce{Sb2S3} and ascribed the observed \SI{0.6}{\electronvolt} Stokes shift to self-trapped excitons, suggesting hopping transport. 
In contrast, \citet{liu2022ultrafast} and \citet{zhang2021suppressing} argued against self-trapping in \ce{Sb2Se3} due to the saturation of fast signal decay with increasing carrier density. 
Considering it is challenging for direct measurements to distinguish whether the photoexcited carriers are intrinsically self-trapped or trapped at defect sites\cite{ramo2007theoretical}, a systematic theoretical study on the carrier transport in \ce{Sb2X3} is necessary. 
The second issue is about the resulting charge carrier mobility. 
Measured mobilities in \ce{Sb2X3} show a large variation\cite{chen2017characterization,liu2016green,zhou2014solution,yuan2016rapid,li2021defect,chalapathi2020influence,black1957electrical}, in part due to different synthesis and characterisation methods.
As such, the intrinsic limits to mobility in \ce{Sb2X3} are unclear and the scattering physics underlying transport is not yet understood. 

In this work, we studied the tendency for polaron trapping and its effect on charge carrier transport in \ce{Sb2X3} by first-principles \ac{DFT} and Boltzmann transport calculations. The electron-lattice interaction in \ce{Sb2X3} was explored through the Fr{\"o}hlich polaron coupling constant and Schultz polaron radius. Modelling of electron and hole polarons in \ce{Sb2X3} indicates the intrinsic formation of large polarons and contrast to recent suggestions of small polarons (i.e.~self-trapped carriers)\cite{yang2019ultrafast,grad2021charge}. The prediction of large polaron formation is further reinforced by the results of carrier transport calculations. 
The isotropically averaged mobilities are higher than \SI{10}{\mob} at room temperature and decrease with increasing temperature for both electrons and holes, further confirming the band-like transport in \ce{Sb2X3}. We find the intrinsic mobility is limited by scattering from polar optical phonons at low and moderate defect concentrations, while at high charged defect concentrations (\textgreater \SI{e18}{\conc}) impurity scattering dominates. We expect our results will enable the design of \ce{Sb2X3} devices with improved efficiencies.

\begin{figure}[ht]
    \centering
    {\includegraphics[width=0.5\textwidth]{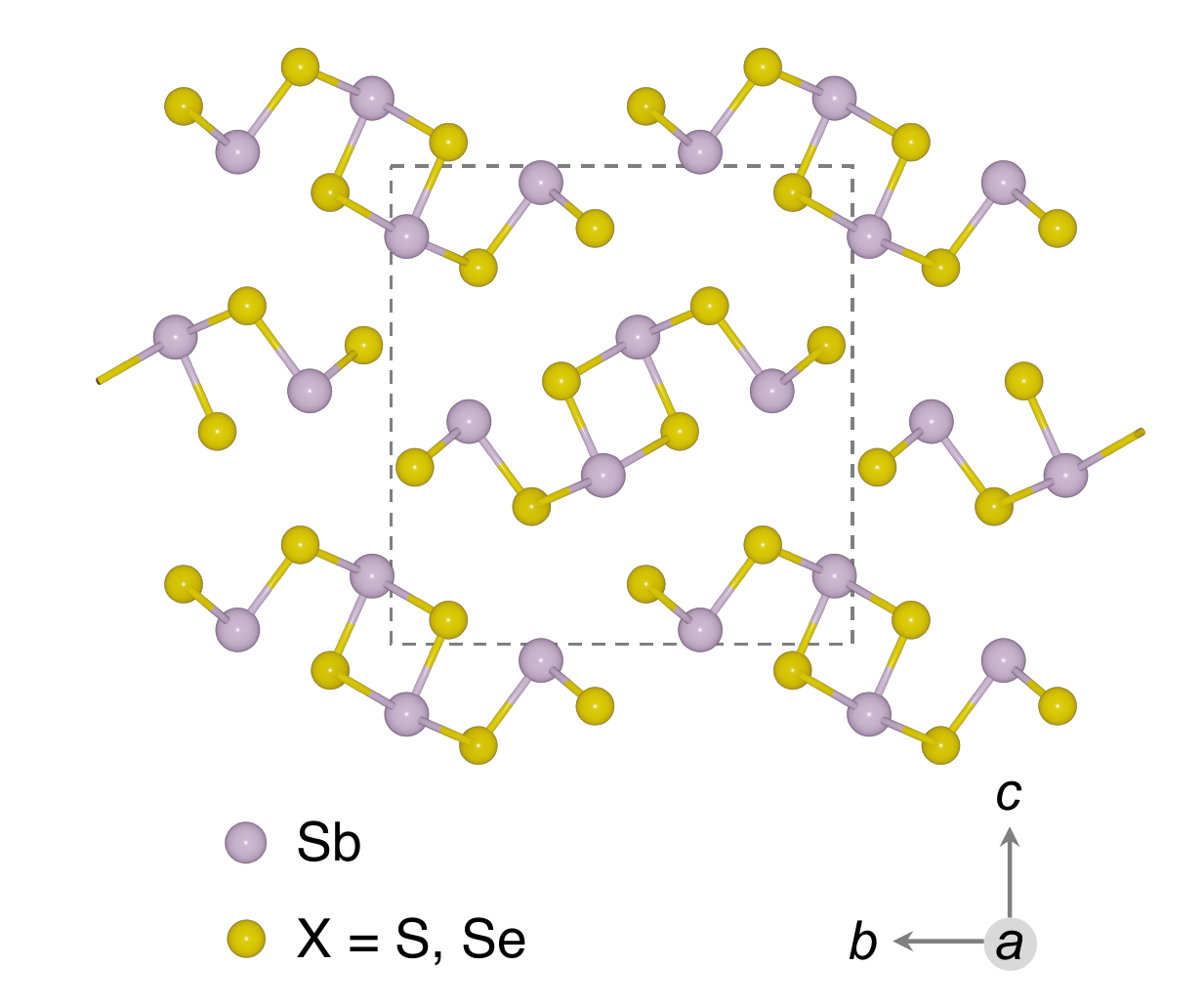}} \\
    \caption{Ground-state crystal structure (\textit{Pnma} space group) of \ce{Sb2X3}. The conventional unit cell is represented by a rectangle.}
    \label{fig_structure}
\end{figure}

\ce{Sb2X3} crystallise in the orthorhombic \textit{Pnma} space group and are comprised of strongly bonded quasi-\ac{1D} [Sb$_4$X$_6$]$_n$ ribbons oriented along the [100] direction (Fig.~\ref{fig_structure}). Ribbon formation is driven by the Sb lone pair with ribbons stacked together by weak interactions\cite{wang2022lone}. According to our previous optimization using the HSE06 hybrid functional and D3 dispersion correction,\cite{wang2022lone} the calculated lattice parameters are 3.80/\SI{3.95}{\angstrom}, 11.20/\SI{11.55}{\angstrom} and 11.39/\SI{11.93}{\angstrom} for \ce{Sb2S3}/\ce{Sb2Se3} along the \textit{a}, \textit{b} and \textit{c} axes, respectively.
\ce{Sb2X3} are indirect band gap semiconductors with calculated indirect/direct band gaps of 1.79/\SI{1.95}{\electronvolt} and 1.42/\SI{1.48}{\electronvolt} for \ce{Sb2S3} and \ce{Sb2Se3}, respectively, which are in reasonable agreement with previous experimental \cite{yesugade1995structural,el1998substrate,versavel2007structural,liu2016green,torane1999preparation,messina2009antimony,lai2012preparation,chen2015optical} and theoretical studies\cite{vadapoo2011self,vadapoo2011electronic,caracas2005first,nasr2011electronic,savory2019complex}. The electronic band structures are shown in Fig.~S1 of the Supplementary Information.
It has been widely suggested that efficient transport can only happen along the ribbons, based on the understanding that \ce{Sb2X3} are \ac{1D} semiconductors\cite{caruso2015excitons,song2017highly,guo2018tunable,yang2018adjusting,gusmao2019antimony}. However, neither the structural dimensionality nor the electronic dimensionality of \ce{Sb2X3} is \ac{1D}.\cite{deringer2015vibrational,wang2022lone} 

\begin{table}[ht]
 \caption{Calculated Fr{\"o}hlich parameter ($\alpha$) and Schultz polaron radius (r$_f$) for electrons (\textit{e}$^-$) and holes (\textit{h}$^+$) in \ce{Sb2S3} and \ce{Sb2Se3} at T = \SI{300}{\kelvin}}
 \label{tab_alpha}
 \begin{tabular*}{\textwidth}{@{\extracolsep{\fill}}cccccc}
\hline
\multicolumn{1}{c}{\multirow{2}{*}{Material}} & \multicolumn{1}{c}{\multirow{2}{*}{}} & \multicolumn{2}{c}{$\alpha$} & \multicolumn{2}{c}{r$_f$ (\AA)} \\ \cline{3-6} 
\multicolumn{1}{c}{} & \multicolumn{1}{c}{} & \multicolumn{1}{c}{\textit{e}$^-$} & \multicolumn{1}{c}{\textit{h}$^+$} & \multicolumn{1}{c}{\textit{e}$^-$} & \multicolumn{1}{c}{\textit{h}$^+$} \\ \hline
\multirow{4}{*}{\ce{Sb2S3}} & avg & 1.6 & 2.0 & 45.5 & 40.4 \\
 & \textit{x} & 1.0 & 1.8 & 57.3 & 43.7 \\
 & \textit{y} & 2.4 & 2.1 & 36.9 & 40.3 \\
 & \textit{z} & 5.7 & 2.5 & 23.7 & 36.4 \\ \hline
\multirow{4}{*}{\ce{Sb2Se3}} & avg & 1.3 & 2.1 & 40.5 & 31.9 \\
 & \textit{x} & 0.8 & 2.0 & 50.9 & 32.4 \\
 & \textit{y} & 2.0 & 1.6 & 32.8 & 36.1 \\
 & \textit{z} & 5.8 & 3.8 & 18.8 & 23.5 \\ \hline
 \end{tabular*}
\end{table}

Charge carriers in crystals are formally described as quasi-particles due to their interaction with the extended structure. In polar semiconductors, the charge carriers and the surrounding lattice deformation form a so-called polaron,\cite{emin2013polarons} which determines the nature of carrier transport.
Polarons can be classified into two types based on the strength of electron-phonon coupling. Stronger coupling leads to larger local lattice distortion which provides the driving force for small polarons to form. Thus, for a small polaron, the lattice deformation is usually confined to one unit cell, and a carrier's motion is typically incoherent with thermally activated hops which lead to low mobility ($\ll$ \SI{1}{\mob}). By contrast, the lattice deformation in a large polaron is usually moderate and spreads over multiple unit cells, resulting in a higher mobility (\textgreater {} \SI{1}{\mob}). 
In polar crystals, the electron-phonon interaction is usually dominated by the coupling of charge carriers to the \ac{LO} phonons, which can be described within the Fr{\"o}hlich model\cite{frohlich1952interaction}. 

We first evaluate the Fr{\"o}hlich interaction by the coupling constant $\alpha$. The calculated $\alpha$ (shown in Table \ref{tab_alpha}) shows an isotropically averaged value of $\sim$2 for both \ce{Sb2S3} and \ce{Sb2Se3}, which falls in the intermediate electron-phonon coupling regime (defined as 0.5 $\lesssim \alpha \lesssim$ 6).\cite{stoneham2001theory} The magnitude of $\alpha$ along the [100] and [010] directions is quite close ($\Delta \alpha$ = 1.2--1.4 and 0.3--0.4 for electrons and holes, respectively), suggesting similar electron-phonon interaction strengths along these two directions. We further estimate the size of polarons in \ce{Sb2X3} by the Schultz polaron radius (r$_f$)\cite{schultz1959slow}. The large values of electron and hole polaron radii (which extend over multiple structural units) indicate the polarons are delocalised in both \ce{Sb2S3} and \ce{Sb2Se3}. The details of parameters used and the procedure for averaging $\alpha$ can be found in Section S2 of the Supplementary Information.

For an alternative assessment, we performed direct first-principles \ac{DFT} calculations to model charge carriers in \ce{Sb2X3}.
There are two challenges for reliable polaron modelling.
The first is the self-interaction error\cite{parr1989w} arising from the approximate form of the exchange-correlation functional which causes electrons to spuriously delocalise\cite{pacchioni2008modeling,pham2020efficient}. This is typically resolved by employing a hybrid functional\cite{finazzi2008excess,deak2011polaronic,di2006electronic} which incorporates a certain amount of exact Fock exchange or by a Hubbard correction (DFT+U)\cite{dudarev1998electron,anisimov1997first}.  
Secondly, the formation of localised polarons is dependent on the initial geometries and wavefunctions. Different methods have been proposed to break the crystal symmetry and promote the formation of localised states. Among them, the bond distortion method and electron attractor method have proved reliable across a range of structures and chemistries\cite{ramo2007theoretical,pham2020efficient,deskins2011distribution,deskins2009localized,shibuya2012systematic,hao2015coexistence,liu2019photocatalytic}. The former involves introducing local perturbations in a supercell in a region where the polaron is expected to localise, while the latter uses a temporarily-substituted atom to attract an electron or a hole, which is then removed and the structure re-relaxes.
In this work, all polaron calculations were performed using the HSE06 hybrid exchange-correlation functional. We attempted to localise electron and hole polarons by adding or removing an electron from a \ce{Sb2X3} supercell using both these distortion methods. The full computational details and workflow are provided in Section S4 (Fig. S3). No energy lowering distortions were found in any case. The electrons and holes always preferred to delocalise rather than localise in both \ce{Sb2S3} and \ce{Sb2Se3} (see Fig. S4 and S5), indicating again that small polarons are unlikely to form intrinsically by self-trapping.
This is also supported by recent experimental evidence that the trap states in \ce{Sb2Se3} are saturated by moderate density photocarriers.\cite{liu2022ultrafast}.

As self-trapping could originate from either self-trapped carriers (i.e. small polarons) or self-trapped excitons, we next consider the possibility of forming self-trapped excitons.
Firstly, the large dielectric constants ($\sim$ 100) and small effective masses ($\sim$ 0.1) in \ce{Sb2X3}\cite{wang2022lone} suggest that the Coulomb interaction is strongly screened and a large exciton radius is favoured. The small experimental exciton binding energies (\SIrange{0.01}{0.05}{\electronvolt} for \ce{Sb2S3} and \SI{0.04}{\electronvolt} for \ce{Sb2Se3})\cite{caruso2015excitons,lawal2018investigation} further indicate weak electron-hole interactions in \ce{Sb2X3}. Additionally, experimental measurements of the imaginary part of the frequency-dependent complex photoconductivity in \ce{Sb2Se3} do not reveal any negative components\cite{wang2019both} that can be a signal of exciton formation. 
Consequently, we conclude that self-trapped excitons in \ce{Sb2X3} are unlikely.

\begin{figure*}[t]
    \centering
    {\includegraphics[width=1.0\textwidth]{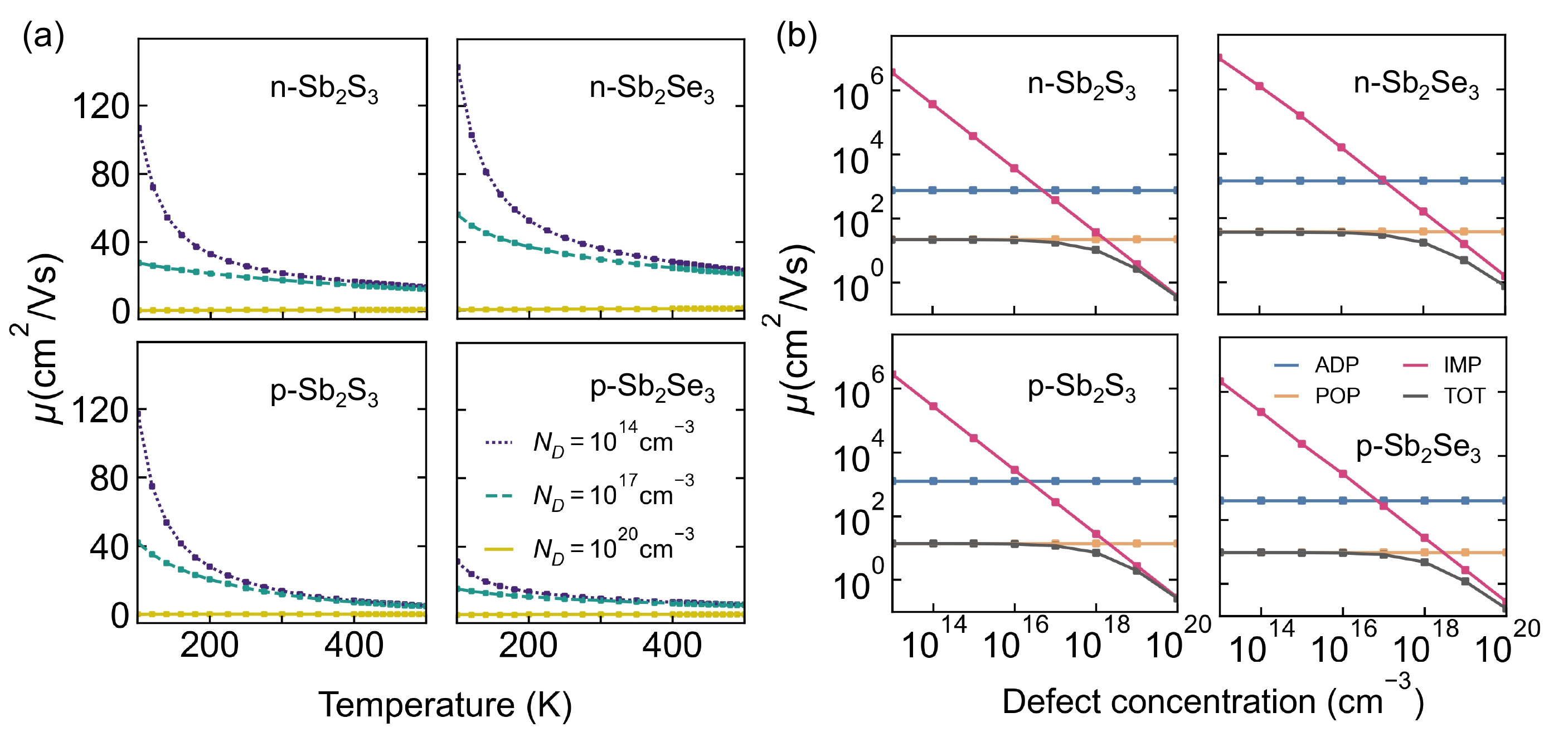}} \\
    \caption{(a) Calculated average mobilities of electrons and holes in \ce{Sb2S3} and \ce{Sb2Se3} as a function of temperature with different defect concentrations. (b) Calculated total and component mobilities as a function of bulk defect concentration at \SI{300}{\kelvin}. ADP, acoustic deformation potential; POP, polar optical phonon; IMP, ionized impurity. $N_D$, defect concentration.}
    \label{fig_mobility_avg}
\end{figure*}

\begin{figure}[t]
    \centering
    {\includegraphics[width=0.5\textwidth]{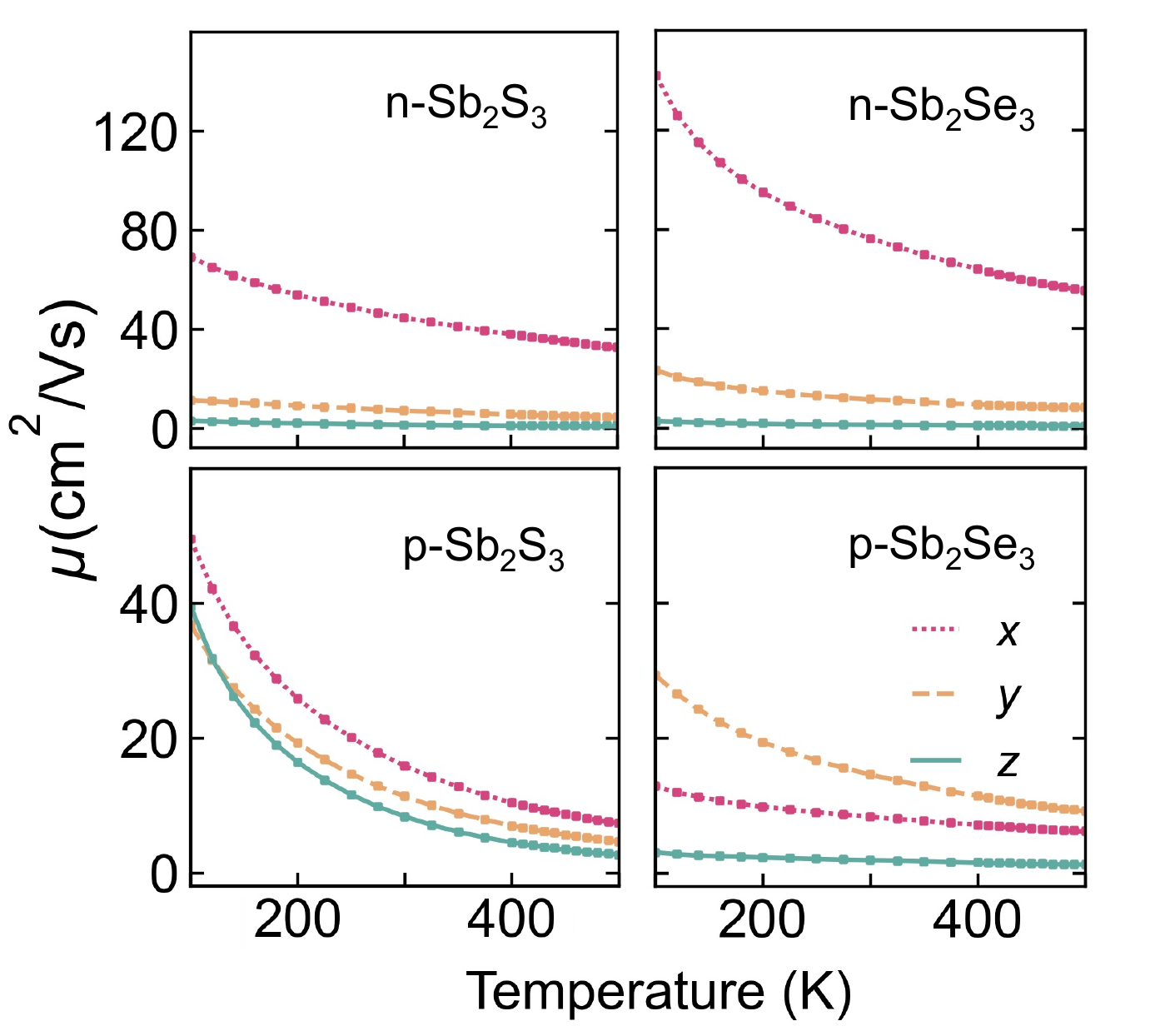}} \\
    \caption{The anisotropic net carrier mobilities including all scattering mechanisms in \ce{Sb2S3} and \ce{Sb2Se3} as a function of temperature with a bulk defect concentration of} \SI{e17}{\conc}.
    \label{fig_mobility_ani}
\end{figure}

\begin{table}[ht]
 \caption{Calculated mobilities of electrons ($\mu_e$) and holes ($\mu_h$) in \ce{Sb2X3} at \SI{300}{\kelvin} under different defect concentrations ($N_D$) and experimental values for comparison. The anisotropy ratio (\textit{a}$_r$) is defined as the ratio of
maximum to minimum mobility}
 \label{tab_mobility}
  \begin{tabular*}{\textwidth}{@{\extracolsep{\fill}}ccccccc}
 \hline
 Material & \multicolumn{2}{c}{} & \multicolumn{3}{c}{Calculated (\si{\mob})} & Experiment (\si{\mob}) \\
\hline
 &  &  & \multicolumn{3}{c}{$N_D$ (cm$^{-3}$)}  &  \\ 
 &  &  & 10$^{14}$ & 10$^{17}$ & 10$^{20}$ &  \\ \cline{4-6} 
\multirow{8}{*}{\ce{Sb2S3}} & \multirow{4}{*}{$\mu_e$} & \textit{x} & 53.90 & 44.72 & 0.96 &  \\
 &  & \textit{y} & 9.60 & 7.13 & 0.07 &  \\
 &  & \textit{z} & 1.88 & 1.35 & 0.01 &  \\
 &  & avg & 21.79 & 17.73 & 0.35 &  \\ 
 &  & \textit{a}$_r$ & 28.67 & 33.13 & 96.00 & \\ \cline{2-7} 
 & \multirow{4}{*}{$\mu_h$} & \textit{x} & 18.58 & 15.90 & 0.38 &  \\
 &  & \textit{y} & 13.53 & 11.33 & 0.19 &  \\
 &  & \textit{z} & 9.34 & 8.35 & 0.22 &  \\
 &  & avg & 13.82 & 11.86 & 0.26 & 6.4-12.8\cite{liu2016green}, 32.2-54.0\cite{chalapathi2020influence} \\ 
  &  & \textit{a}$_r$ & 1.99 & 1.90 & 2.00 & \\ \hline
\multirow{8}{*}{\ce{Sb2Se3}} & \multirow{4}{*}{$\mu_e$} & \textit{x} & 89.97 & 76.38 & 1.96 &  \\
 &  & \textit{y} & 16.74 & 11.65 & 0.11 &  \\
 &  & \textit{z} & 1.94 & 1.41 & 0.01 &  \\
 &  & avg & 36.22 & 29.81 & 0.70 & 15\cite{black1957electrical} \\ 
  &  & \textit{a}$_r$ & 46.38 & 54.17 & 196.00 & \\ \cline{2-7}
 & \multirow{4}{*}{$\mu_h$} & \textit{x} & 9.50 & 8.38 & 0.17 & 2.59\cite{chen_characterization_2017} \\
 &  & \textit{y} & 16.95 & 14.63 & 0.25 & 1.17\cite{chen_characterization_2017} \\
 &  & \textit{z} & 2.22 & 1.95 & 0.06 & 0.69\cite{chen_characterization_2017} \\
 &  & avg & 9.55 & 8.32 & 0.16 & 5.1\cite{zhou2014solution}, 3.7-21.88\cite{yuan2016rapid}, 45\cite{black1957electrical} \\ 
  &  & \textit{a}$_r$ & 7.64 & 7.50 & 4.17 & \\ \hline
\end{tabular*}
\end{table}

To further understand the nature of transport in \ce{Sb2X3} the first-principles carrier mobility\cite{ganose2021efficient} was calculated. Both \textit{n}-type and \textit{p}-type doping were investigated, with calculations including scattering from \ac{IMP}, \ac{ADP} and \ac{POP}.
Piezoelectric scattering was not considered due to the centrosymmetric crystal structure.
The isotropically averaged mobilities are reasonably high at room temperature (T = \SI{300}{\kelvin}) for both electrons ($\sim$\SI{40}{\mob}) and holes ($\sim$\SI{15}{\mob}), at low and moderate defect concentrations ($<$\SI{1e18}{\conc}), indicating band-like transport (Fig.~\ref{fig_mobility_avg}a).
The hole mobilities are a little lower than the electron mobilities in both \ce{Sb2S3} and \ce{Sb2Se3}, suggesting that \textit{n}-type doping could be beneficial for carrier collection in photovoltaic devices.
This is in contrast to experimental measurements that have indicated higher mobility for \textit{p}-type \ce{Sb2Se3},\cite{black1957electrical} however, this may be related to the doping asymmetry in these materials.
The intrinsic mobility is limited by Fr{\"o}hlich-type polar optical phonon scattering suggesting that large polarons are responsible for the transport behaviour (Fig.~\ref{fig_mobility_avg}b).
We note that large deformation potentials have been suggested as the origin of self-trapping in the bismuth double perovskites\cite{wu2021strong}.
However, in \ce{Sb2X3}, acoustic deformation potential scattering is weak (due to small deformation potentials $<$ \SI{6}{\electronvolt}), similar to that seen in the hybrid halide perovskites\cite{wright2016electron,lu2017piezoelectric}, indicating self-trapping is unlikely to occur via coupling with acoustic vibrations.

The scattering from ionized impurities increases with the defect concentration. 
At concentrations around \SI{e18}{\conc}, \ac{IMP} and \ac{POP} scattering are roughly the same strength and cause the mobility to reduce by a factor of a half (Fig.~\ref{fig_mobility_avg}b).
At higher defect concentrations transport is entirely dominated by ionized impurities.
Our results indicate that careful control of defect concentrations are essential for preventing degradation of device efficiencies.
This agrees well with previous experimental reports that the defect density is crucial to the carrier transport in \ce{Sb2X3}, whereby bulk defect densities above \SI{e15}{\conc} led to significant degradation in conversion efficiency \cite{islam2020two,li2020simulation,khadir2022performance}. 
Furthermore, considering that most experimental mobility measurements in \ce{Sb2X3} were obtained from thin films where grain boundary scattering will further lower the mobility, we also tested the inclusion of mean free path scattering. According to our results (Fig. S2 and Table S3), the mobilities in \ce{Sb2X3} are not significantly affected by grain boundary scattering even with grain sizes down to \SI{10}{\nm}, much smaller than the domain sizes typically seen in experiments\cite{rijal2021influence,maghraoui2010structural,perales2008optical,lokhande2001novel}.
Accordingly, our results suggest that grain boundary scattering is unlikely to be a dominant source of scattering in \ce{Sb2X3} thin films, in agreement with previous studies\cite{gonzalez2022deciphering}.

The anisotropy of mobility was also considered. As shown in Table \ref{tab_mobility} and Fig.~\ref{fig_mobility_ani}, our calculated mobilities are in reasonable agreement with the range of measured values.
For electron transport, there is considerable anisotropy with the [100] direction showing roughly 6 times the mobility of the [010] direction and over 30 times the mobility of the [001] direction in both \ce{Sb2S3} and \ce{Sb2Se3}.
For holes in \ce{Sb2S3}, there is a high mobility in the (001) plane where the transport is roughly isotropic and approximately twice that of the [001] direction.
For holes in \ce{Sb2Se3}, the picture is slightly altered with the highest mobility seen along [010], roughly 2 times the mobility along [100] and 8 times the mobility along [001].
The anisotropy in mobility follows the anisotropy in the calculated effective masses and the Fermi-surface dimensionality\cite{wang2022lone}.
Thus, as the electron mobilities are higher and more anisotropic than the hole mobilities, control of the grain orientation is necessary to achieve more efficient electronic transport in devices, which can be realised by strategies such as seed screening\cite{li2019orientation} and quasi-epitaxial growth\cite{wang2017stable}.
Despite the anisotropic behaviour, even at moderate defect concentrations the electron and hole mobilities are still reasonably high ($>$\SI{10}{\mob}) in at least two directions.
The common description of \ce{Sb2X3} as a \ac{1D} semiconductors\cite{zhou2015thin,liang2020crystallographic} oversimplifies the nature of transport.
Accordingly, it may be possible to obtain high mobility thin films, even when the grains are not fully aligned along the direction of the quasi-1D ribbons.

Conclusively, our results show no evidence for carrier self-trapping in \ce{Sb2X3}. While self-trapping has been proposed based on several experimental observations,\cite{yang2019ultrafast,tao2022coupled} 
large polaron carriers do not contradict these observations: 
\textrm{i}) a Stokes shift of \SI{0.6}{\electronvolt} and broad \ac{PL};  
\textrm{ii}) picosecond carrier decay kinetics; 
\textrm{iii}) absence of photoexcited carrier density saturation (up to \SI{e20}{\conc}); and 
\textrm{iv}) polarized light emission in \ce{Sb2X3} single crystals. 
Firstly, a large Stokes shift and broad PL is found in many chalcogenide semiconductors, especially those with deep defect levels such as \ce{Sb2X3}.
Secondly, the timescale for carrier decay due to self-trapping is typically sub-picosecond or several picoseconds,
while a timescale of tens of picoseconds is found in \ac{TA} measurements of \ce{Sb2X3}.\cite{yang2019ultrafast,tao2022coupled}
We note that the understanding of TA kinetics in indirect band gap semiconductors is still evolving.
Thirdly, the TA signal persists to high carrier densities (excitation power), which could also be explained by photoinduced absorption or a large trap density $>$ \SI{e20}{\conc}.\cite{zhang2021suppressing}
Finally, polarized light emission is found in many semiconductors and is connected to the crystal and defect structure. 
Therefore, we find no evidence that directly supports intrinsic carrier self-trapping in these materials. 

In summary, we investigated the nature of charge carriers in \ce{Sb2X3} semiconductors. 
Our results strongly suggest that self-trapping (i.e. the formation of small polarons) is unlikely to occur and that instead charge transport involves large polarons with moderate mobility. 
In particular, we found:
i) intermediate Fr{\"o}hlich coupling constants ($\sim$2); ii) large Schultz polaron radii ($\sim$\SI{40}{\angstrom}); 
iii) the absence of electron or hole polaron formation in DFT calculations using the bond distortion and electron attractor methods; and iv) carrier mobilities \textgreater \SI{10}{\mob} at room temperature for both electrons and holes (in agreement with experiments).
We conclude that there is no theoretical evidence for small polaron formation in pristine \ce{Sb2X3} and self-trapping is unlikely to be the origin of the low open-circuit voltages in \ce{Sb2X3} devices as reported in previous studies\cite{yang2019ultrafast,grad2021charge}.
Accordingly, the low photovoltages may not be a bulk property of these materials and could be surmountable with improved fabrication and processing conditions to engineer the defect and interfacial properties of devices.

\section{Methods}
The Fr{\"o}hlich polaron properties were solved using the open-source package \textsc{PolaronMobility}\cite{Frost2017}. 
The first-principles carrier scattering rates and resulting mobilities were calculated using \textsc{AMSET}\cite{ganose2021efficient}. 
The set of materials parameters used for these predictions are provided in Table S1--S2, S4--S6.
The crystal structure was plotted using \textsc{Blender}\cite{blender} and \textsc{Beautiful Atoms}\cite{Beautiful_Atoms2022}.

All of the underlying electronic structure calculations were performed based on Kohn-Sham density-functional theory\cite{kohn1965self,dreizler1990density} as implemented in \ac{VASP}\cite{kresse1996efficient}. The projector augmented-wave (PAW) method\cite{kresse1999ultrasoft} was employed with a plane-wave energy cutoff of \SI{400}{\electronvolt}. All calculations were carried out using the Heyd-Scuseria-Ernzerhof hybrid functional (HSE06)\cite{heyd2003hybrid,krukau2006influence} with the D3 dispersion correction\cite{grimme2004accurate}, which have been proved to be able to well describe the structural and electronic properties in \ce{Sb2X3}\cite{wang2022lone}. The atomic positions were optimised until the Hellman-Feynman forces on each atom were below \SI{0.0005}{\electronvolt\per\angstrom} for unit cells and \SI{0.01}{\electronvolt\per\angstrom} for 3$\times$1$\times$1 supercells. 
The energy convergence criterion was set to \SI{e-6}{\electronvolt}. $\varGamma$-centered \textit{k}-point meshes were set to 7$\times$2$\times$2 and 2$\times$2$\times$2 for geometry optimisation with primitive unit cells and supercells, respectively. For uniform band structure calculations which were used as inputs for AMSET, a denser \textit{k}-point mesh of 19$\times$10$\times$10 was used which is consistent with our previous calculations of carrier effective masses\cite{wang2022lone}. Detailed settings and convergence data are presented in Section S6 (Fig. S6 and Table S4--S6).

\section*{Acknowledgements}
X.W. thanks Jarvist M. Frost, Yuchen Fu and Ye Yang for valuable discussions.
We are grateful to the UK Materials and Molecular Modelling Hub for computational resources, which is partially funded by EPSRC (EP/P020194/1 and EP/T022213/1). X.W. acknowledges Imperial College London for a President's PhD Scholarship. A.M.G. was supported by EPSRC Fellowship EP/T033231/1. S.R.K. acknowledges the EPSRC Centre for Doctoral Training in the Advanced Characterisation of Materials (CDT-ACM)(EP/S023259/1) for a PhD studentship. 

\begin{suppinfo}
Electronic band structures; formulas and input data of Fr{\"o}hlich polaron coupling constant and Schultz polaron radius; effect of grain boundary scattering; workflow of localising a polaron; mobility parameters.
Data produced during this work is freely available at: \url{ https://dx.doi.org/10.17172/NOMAD/2022.08.11-2}.
\end{suppinfo}

\section*{Author Contributions}
The author contributions have been defined following the CRediT system.
X.W.: Conceptualization, Investigation, Formal analysis, Methodology, Visualization, Writing – original draft. 
A.M.G.: Methodology, Supervision, Writing – review \& editing. 
S.R.K.: Methodology, Writing – review \& editing. 
A.W.: Conceptualization, Methodology, Supervision, Writing – review \& editing.


\bibliography{References}

\end{document}


\section*{S1. Electronic band structures}
\setlength{\parindent}{0pt}
\begin{figure}[h]
    \centering
    {\includegraphics[width=0.8\textwidth]{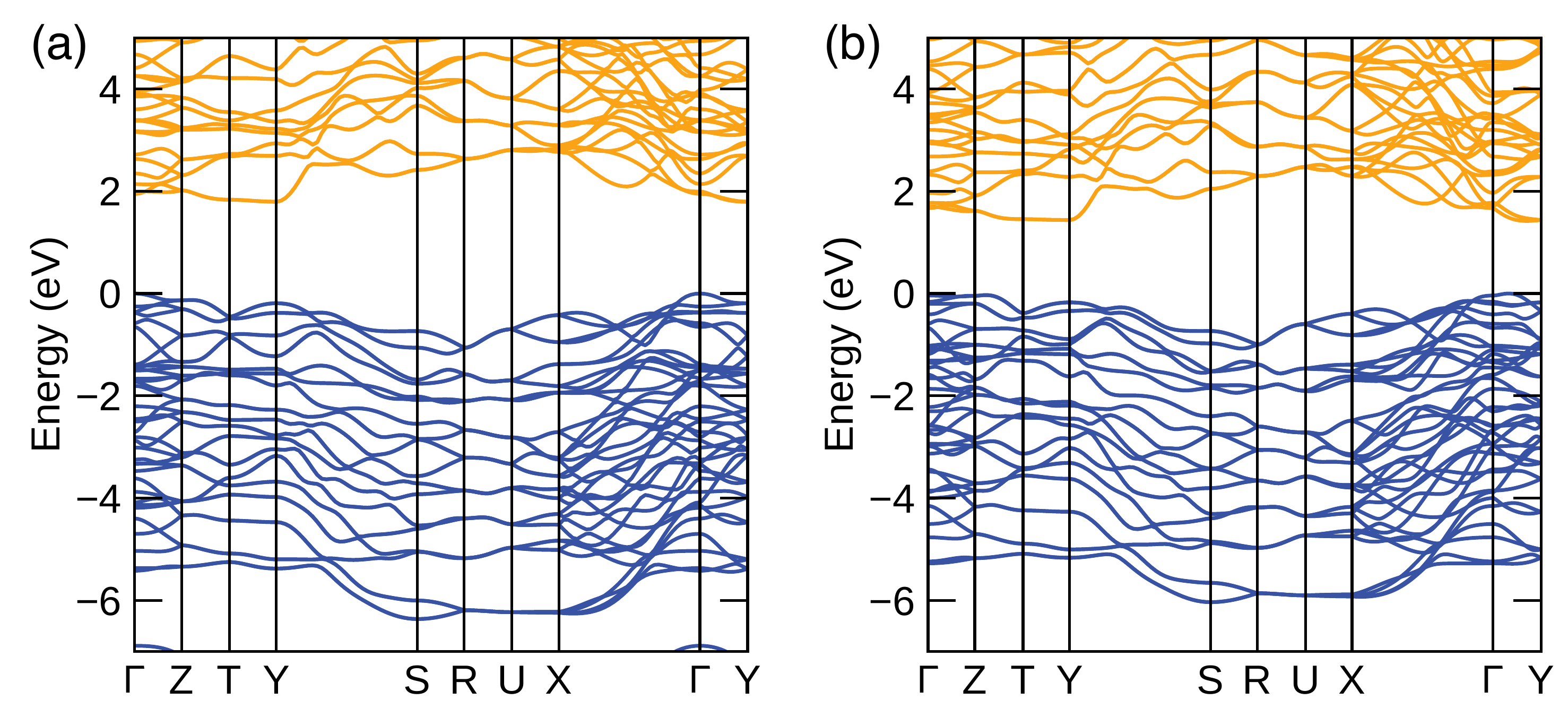}} \\
    \caption{Electronic band structures of (a) \ce{Sb2S3} and (b) \ce{Sb2Se3}.}
    \label{fig_band}
\end{figure}

\section*{S2. Fr{\"o}hlich polaron coupling constant and Schultz polaron radius}
\begin{table}[h]
 \label{tab_alpha_p}
 \caption{Parameters used to calculate Fr{\"o}hlich polaron coupling constant $\alpha$. The effective phonon frequency ($\omega$) is in THz}
\begin{tabular*}{\textwidth}{@{\extracolsep{\fill}}ccccccc}
\hline
\multirow{2}{*}{Material} & \multirow{2}{*}{} & \multirow{2}{*}{\textit{\textepsilon}$_{\infty}$} & \multirow{2}{*}{\textit{\textepsilon}$_0$} & \multirow{2}{*}{\textit{$\omega$}} & \multicolumn{2}{c}{\textit{m}$^*$} \\ \cline{6-7} 
 &  &  &  &  & e & h \\ \hline
\multirow{4}{*}{\ce{Sb2S3}} & avg & \multirow{4}{*}{10.26} & \multirow{4}{*}{68.76} & \multirow{4}{*}{3.49} & 0.40 & 0.64 \\
 & \textit{x} &  &  &  & 0.16 & 0.47 \\
 & \textit{y} &  &  &  & 0.92 & 0.65 \\
 & \textit{z} &  &  &  & 5 & 0.97 \\ \hline
\multirow{4}{*}{\ce{Sb2Se3}} & avg & \multirow{4}{*}{13.52} & \multirow{4}{*}{76.27} & \multirow{4}{*}{2.57} & 0.35 & 0.90 \\
 & \textit{x} &  &  &  & 0.14 & 0.85 \\
 & \textit{y} &  &  &  & 0.81 & 0.55 \\
 & \textit{z} &  &  &  & 7 & 3 \\ \hline
\end{tabular*}
\end{table}
The long-range electron-longitudinal optical phonon coupling can be expressed by the dimensionless Fr{\"o}hlich polaron coupling constant $\alpha$\cite{frohlich1952interaction}
\begin{equation}
\alpha=\frac{e^2}{\hbar}(\frac{1}{\varepsilon_\infty}-\frac{1}{\varepsilon_0})\sqrt{\frac{m^*}{2\hbar\omega}},
\end{equation}
where \textit{\textepsilon}$_{\infty}$ and \textit{\textepsilon}$_0$ are the high-frequency and static dielectric constants, respectively, \textit{m}$^*$ is the effective mass and $\omega$ is the effective phonon frequency. The effective mass and effective frequency were calculated using the AMSET package\cite{ganose2021efficient}.
The isotropic $\alpha$ was obtained using the harmonic mean of the effective masses and the arithmetic average of the dielectric constants. The anisotropic $\alpha$ was calculated using the anisotropic (direction-dependent) effective masses, consistent with previous work\cite{guster2021frohlich}.
~\\
~\\
~\\
~\\
~\\
\begin{table}[h]
 \label{tab_rf}
 \caption{Parameters used to calculate Schultz polaron radius (r$_f$, \AA)}
\begin{tabular*}{\textwidth}{@{\extracolsep{\fill}}cccccccccc}
\hline
\multirow{2}{*}{Material} & \multirow{2}{*}{} & \multicolumn{2}{c}{$\alpha$} & \multicolumn{2}{c}{\textit{v}} & \multicolumn{2}{c}{\textit{w}} & \multicolumn{2}{c}{\textit{m}$^*_\textrm{P}$} \\ \cline{3-10} 
& & \textit{e}$^-$  & \textit{h}$^+$  & \textit{e}$^-$  & \textit{h}$^+$ & \textit{e}$^-$ & \textit{h}$^+$ & \textit{e}$^-$  & \textit{h}$^+$\\ \hline
\multirow{4}{*}{\ce{Sb2S3}} & avg &1.6  &2.0 &12.70  &12.98 &11.16  & 10.99& 0.52 & 0.89 \\
 & \textit{x} & 1.0 &1.8 &12.34 &12.79   &11.39 &11.11  & 0.19 & 0.62 \\
 & \textit{y} &2.4 &2.1 &13.25  &12.99 &10.82 &10.98  & 1.38 & 0.91 \\
 & \textit{z} &5.7 &2.5 &16.07  &13.30 &9.32 &10.79 & 14.88 & 1.47 \\ \hline
\multirow{4}{*}{\ce{Sb2Se3}} & avg &1.3  &2.1  &16.72  & 17.29 & 15.32 & 14.96 & 0.42 & 1.20 \\
 & \textit{x} & 0.8 &2.0&16.40 &17.24 &15.53 & 14.99 & 0.16 & 1.12 \\
 & \textit{y} &2.0 &1.6 &17.21  &16.95 &15.01 &15.17  & 1.06 & 0.69 \\
 & \textit{z} &5.8 &3.8 &20.68 &18.68 &13.05 &14.12  & 17.59 & 5.25 \\ \hline
 \end{tabular*}
\end{table}

Schultz polaron radius is defined as\cite{schultz1959slow}
\begin{equation}
r_f=\sqrt{\frac{3}{2\mu\textit{v}}},
\end{equation}

\begin{equation}
\mu=\frac{v^2-w^2}{v^2},
\end{equation}

where \textit{v} and \textit{w} are Feynman-model variational parameters which specify the polaron state. They are solved variationally by the Feyman polaron model using Fr{\"o}hlich polaron coupling constant $\alpha$ as an input. $\mu$ is the reduced effective mass.

\newpage

\section*{S3. Effect of grain boundary scattering}
\begin{table}[ht]
 \caption{Calculated mobilities of electrons ($\mu_e$) and holes ($\mu_h$) in \ce{Sb2X3} at \SI{300}{\kelvin} with and without grain boundary scattering. The anisotropy ratio (\textit{a}$_r$) is defined as the ratio of
maximum to minimum mobility}
 \label{tab_mfp}
  \begin{tabular*}{\textwidth}{@{\extracolsep{\fill}}cccccc}
 \hline
 Material & \multicolumn{2}{c}{} & \multicolumn{3}{c}{Calculated mobility (\si{\mob})}  \\

\hline
 &  &  & \multicolumn{3}{c}{Mean free path (nm)}  \\ 
 &  &  &  - & 100 & 10  \\ \cline{4-6} 
\multirow{10}{*}{\ce{Sb2S3}} & \multirow{5}{*}{$\mu_e$} & \textit{x} & 44.72 &43.98  &38.57   \\
 &  & \textit{y} & 7.13 &7.07  & 6.55  \\
 &  & \textit{z} & 1.35 &1.34  &1.25   \\
 &  & avg & 17.73 &17.46  &15.45   \\ 
 &  & \textit{a}$_r$ & 33.13 & 32.82 &30.86 \\ \cline{2-6} 
 & \multirow{4}{*}{$\mu_h$} & \textit{x} & 15.90 & 15.77 &14.71 \\
 &  & \textit{y} & 11.33 & 11.25 & 10.58 \\
 &  & \textit{z} & 8.35 &8.29  &7.82  \\
 &  & avg & 11.86 &11.77 & 11.04 \\ 
  &  & \textit{a}$_r$ & 1.90 &1.90  &1.88 \\ \hline
\multirow{10}{*}{\ce{Sb2Se3}} & \multirow{5}{*}{$\mu_e$} & \textit{x} & 76.38&74.56  &62.14  \\
 &  & \textit{y} & 11.65&11.46  &10.07 \\
 &  & \textit{z} & 1.41 &1.39 &1.23 \\
 &  & avg &  29.81 &29.13 & 24.48 \\ 
  &  & \textit{a}$_r$ & 54.17 & 53.64  &50.52  \\ \cline{2-6}
 & \multirow{4}{*}{$\mu_h$} & \textit{x} & 8.38 &8.29  &7.57  \\
 &  & \textit{y} & 14.63 &14.41  &12.78 \\
 &  & \textit{z} & 1.95& 1.93 &1.81  \\
 &  & avg & 8.32&8.21  &7.38  \\ 
  &  & \textit{a}$_r$ & 7.50 & 7.47 &7.06  \\ \hline
\end{tabular*}
\end{table}

\begin{figure}[ht]
    \centering
    {\includegraphics[width=1.0\textwidth]{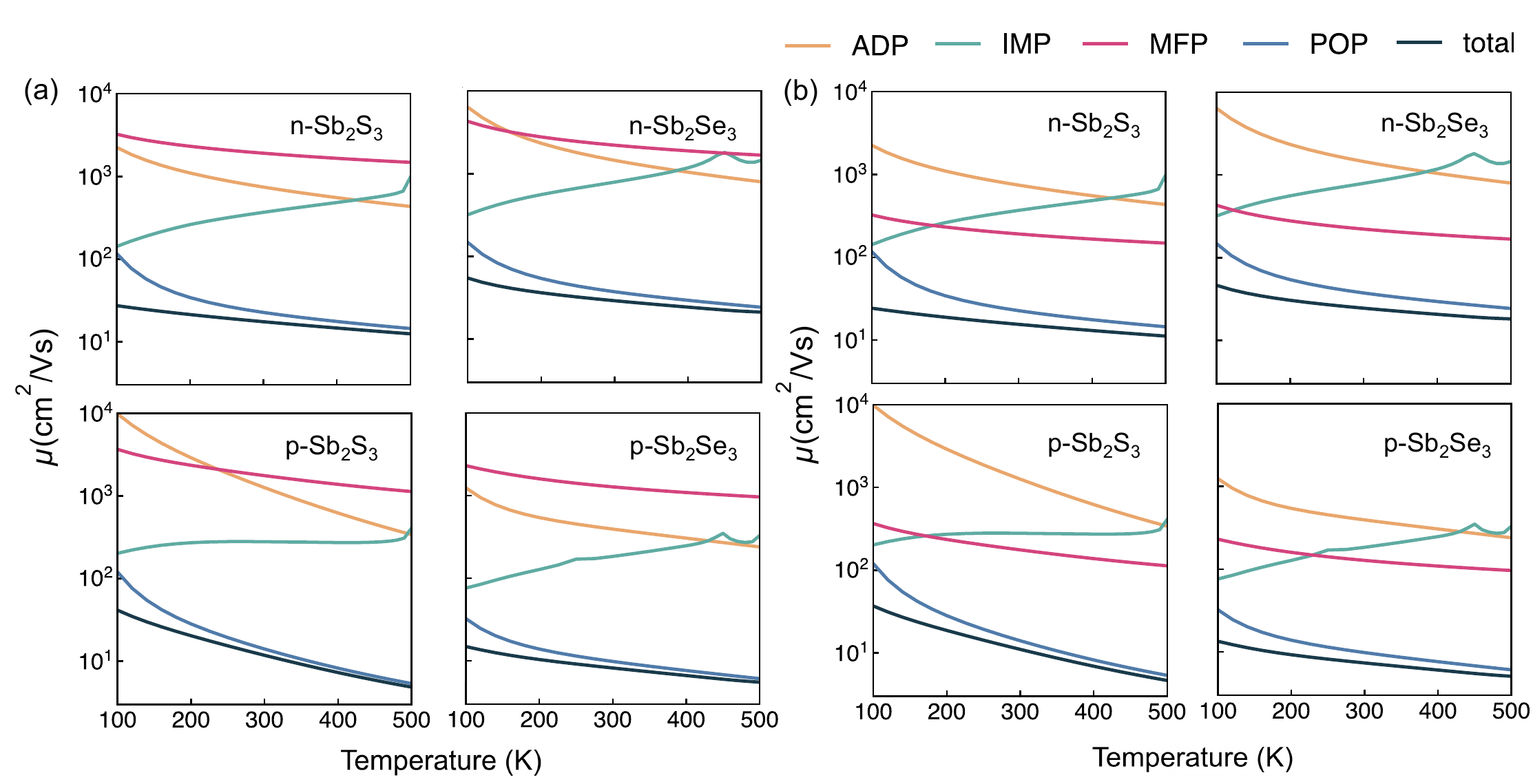}} \\
    \caption{Calculated component and total mobilities with mean free path of (a) 100 nm and (b) 10 nm as a function of temperature.}
    \label{fig_mfp}
\end{figure}

The effect of grain boundary scattering on the mobility in \ce{Sb2X3} was evaluated by incorporating an average grain size using the AMSET package\cite{ganose2021efficient}. The grain boundary scattering lifetime is set to $v_g/L$, where $v_g$ is the group velocity and \textit{L} is the mean free path. In this work, the mean free path of \SI{10} and \SI{100}{\nm} were tested. The carrier concentration and defect concentration were assumed to be \SI{e13}{\conc} and \SI{e17}{\conc}, respectively.
~\\
~\\
According to our results (Table \ref{tab_mfp} and Fig. \ref{fig_mfp}), at temperatures between 100 and \SI{500}{\kelvin}, the total mobility is not limited by the grain boundary scattering. The anisotropic values at room temperature are shown in Table \ref{tab_mfp}. After considering the grain boundary scattering, the values of anisotropy ratio change slightly and the most favourable directions for carrier transport remain the same for both \ce{Sb2S3} and \ce{Sb2Se3}.
~\\

\newpage

\section*{S4. Workflow of localising a polaron in \ce{Sb2X3}}

\begin{figure}[ht]
    \centering
    {\includegraphics[width=1.0\textwidth]{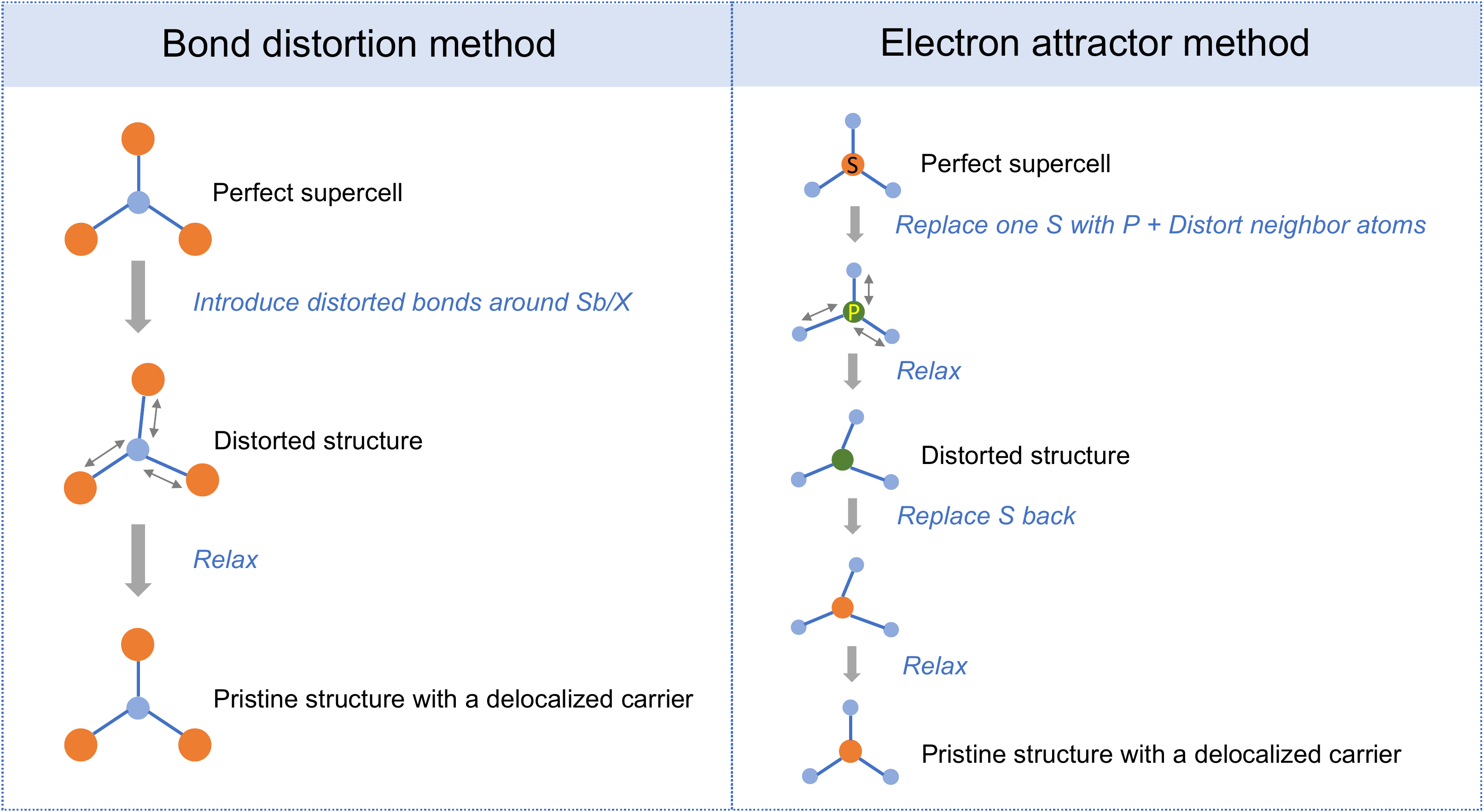}} \\
    \caption{The workflow of bond distortion method and electron attractor method.}
    \label{fig_wf}
\end{figure}

\begin{figure}[ht]
    \centering
    {\includegraphics[width=1.0\textwidth]{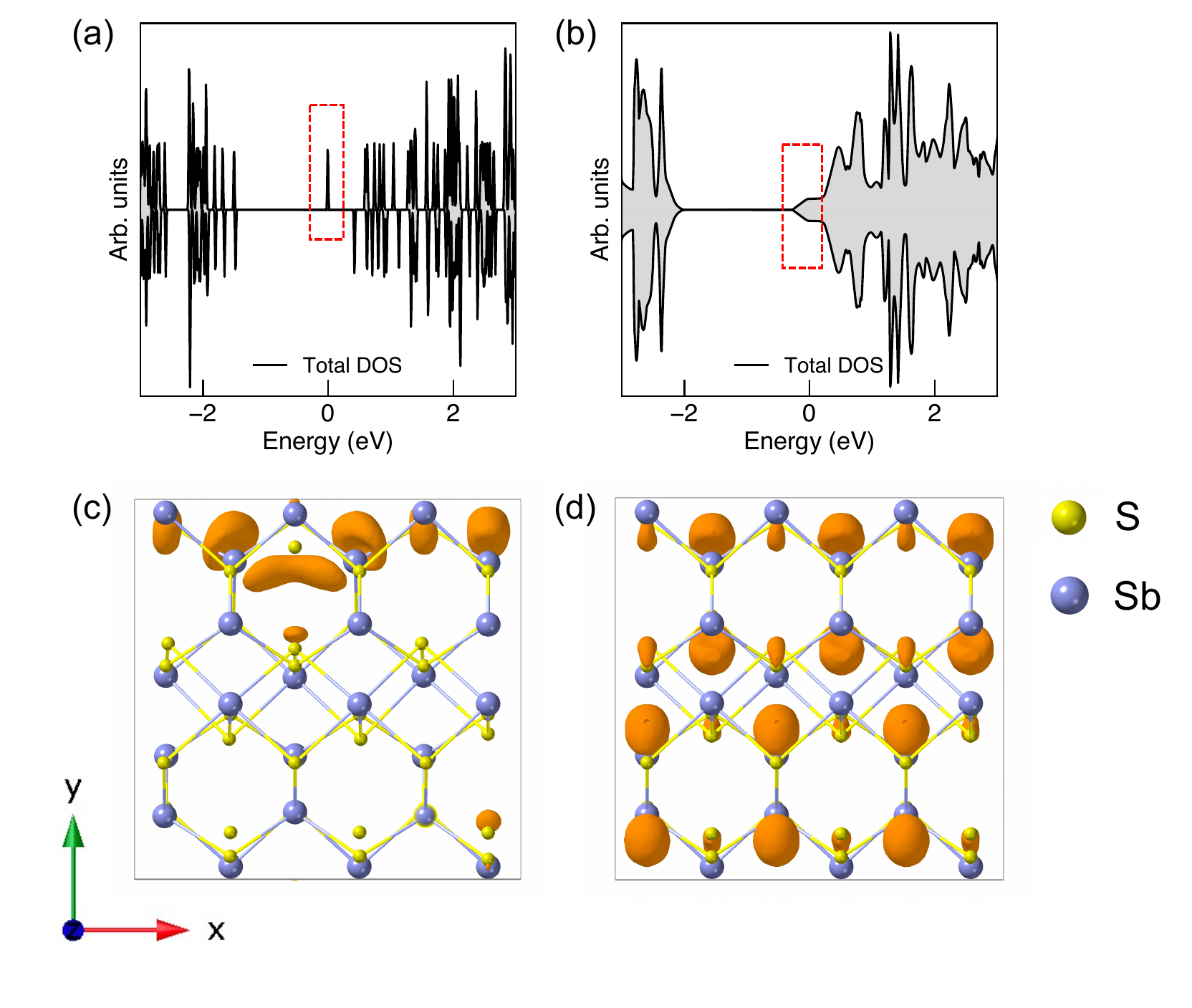}} \\
    \caption{(a-b) Density of states and (c-d) partial charge densities of conduction band maximum for electron polarons in \ce{Sb2S3} using a \textit{k}-point mesh of (a) and (c) 1$\times$1$\times$1, (b) and (d) 2$\times$2$\times$2 to relax structures. The partial charge densities were plotted by specifying the energy range of the conduction band maxima, which are represented by red dashed rectangles. The isosurface value for partial charge densities is set to 0.01 \textit{e}/\AA$^3$.}
    \label{fig_k}
\end{figure}

We attempted to localise an electron or a hole in \ce{Sb2S3} and \ce{Sb2Se3} by the bond distortion method and electron attractor method (Fig. \ref{fig_wf}). A 3$\times$1$\times$1 supercell (with the dimension of 11.40$\times$11.20$\times$\SI{11.39}{\cubic\angstrom} and 11.85$\times$11.55$\times$\SI{11.93}{\cubic\angstrom} for \ce{Sb2S3}  and \ce{Sb2Se3}, respectively) was constructed, which is sufficient to model small polarons\cite{sun2017disentangling,ding2014computational,castleton2019benchmarking}. In each system, one electron per supercell was added or removed to introduce an electron or a hole.
~\\
~\\
We first applied the bond distortion method to introduce distortions around one designated atom (Sb for adding an electron and S/Se for adding a hole) for each non-equivalent Sb and S/Se. These are implemented by the ShakeNBreak package\cite{mosquera2021search,mosquera2022identifying}. Different distortions of 20\%, 30\% and 40\% with both compression and stretching were considered. However, after structural optimisation, all lowest-energy structures relaxed to perfect configurations.
~\\
\newpage
We further combined the bond distortion method with the electron attractor method to confirm the formation of hole polarons in \ce{Sb2S3}. The electron attractor method refers to attracting electrons or holes to a particular atomic site by replacing one certain atom. Phosphorous has stronger attraction to holes than sulfur as it contains fewer protons and has a lower electronegativity. Here, we used one P to replace one S in a supercell, introduced some local distortions around the P atom and add small random displacements to all atoms to break the symmetry in the initial structures. Three non-equivalent S sites were considered, and a range of distortions of both compression and stretching between 0\% and 60\% with 10\% as an interval were tested. The number of electrons were kept the same as the neutral replaced system, suggesting one extra hole in \ce{Sb2S3}. The structures with the substituted atom and local distortions were fully relaxed. Finally, for each non-equivalent S case, we used the lowest-energy structures among different distortions, replaced back the S atom and relaxed the configuration again. Nevertheless, all structures went back to perfect configurations, indicating that the localised polarons are unlikely to form.
~\\
~\\
Nevertheless, we note that using a \textit{k}-point mesh of 1$\times$1$\times$1 to do geometry relaxation could lead to localised solution in some cases (Fig. \ref{fig_k}a). While after converging it with denser \textit{k}-point mesh of 2$\times$2$\times$2, we finally get delocalised polarons (Fig. \ref{fig_k}b).
\newpage

\newpage

\section*{S5. Partial charge densities of electron and hole polarons}
\begin{figure}[ht]
    \centering
    {\includegraphics[width=1.0\textwidth]{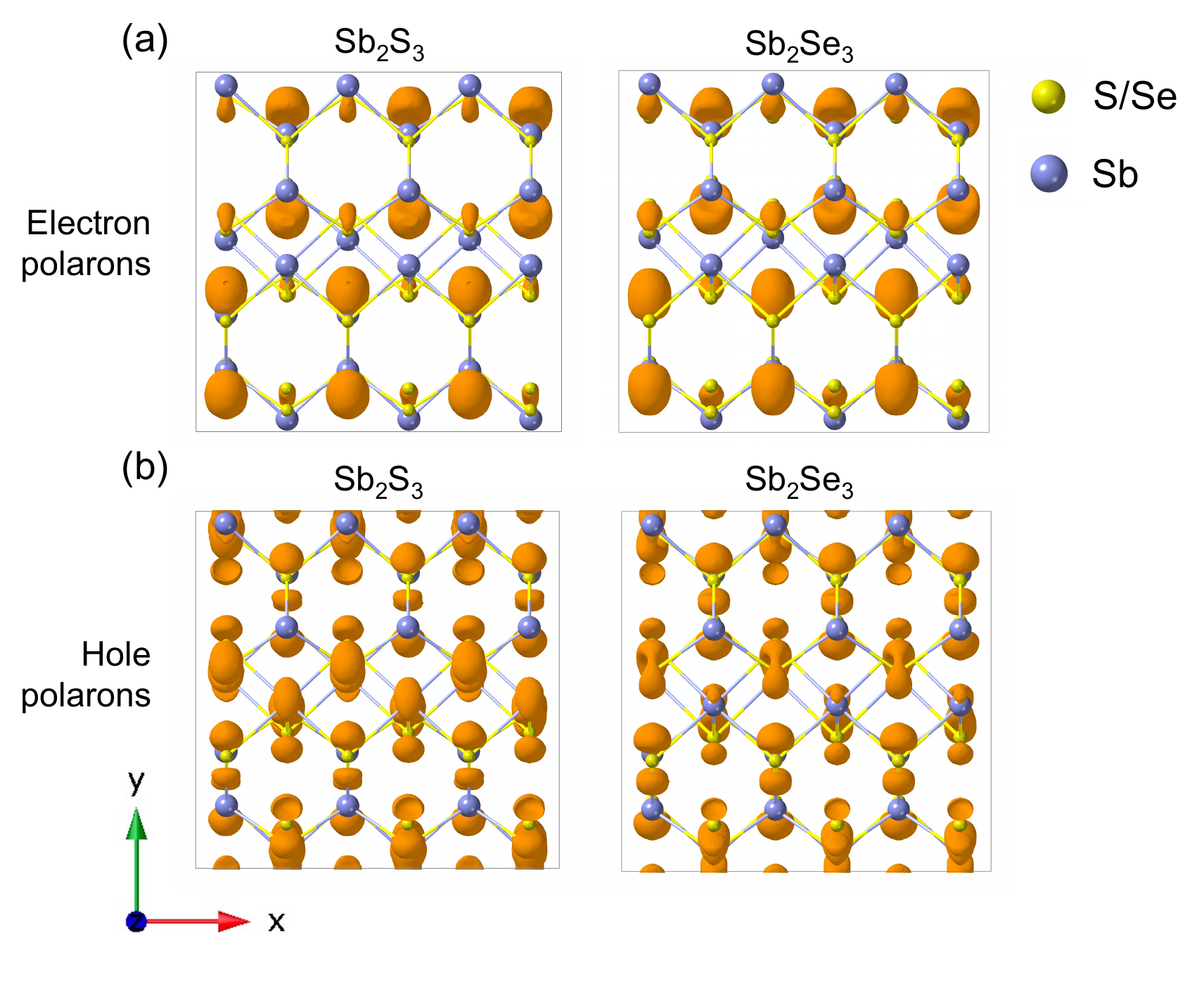}} \\
    \caption{Partial charge densities of (a) conduction band maximum for electron polarons and (b) valence band maximum for hole polarons in \ce{Sb2S3} and \ce{Sb2Se3}. The isosurface values for partial charge densities are set to 0.01 and 0.005 \textit{e}/\AA$^3$ for electron and hole polarons, respectively.}
    \label{fig_pc}
\end{figure}

Partial charge densities of the valence band maximum (VBM) for hole polarons and conduction band maximum (CBM) for electron polarons in \ce{Sb2X3} are shown in Fig. \ref{fig_pc}, which are delocalised in all cases. 

\section*{S6. Parameters used to calculate mobilities in \ce{Sb2X3}}
\begin{figure}[ht]
    \centering
    {\includegraphics[width=0.8\textwidth]{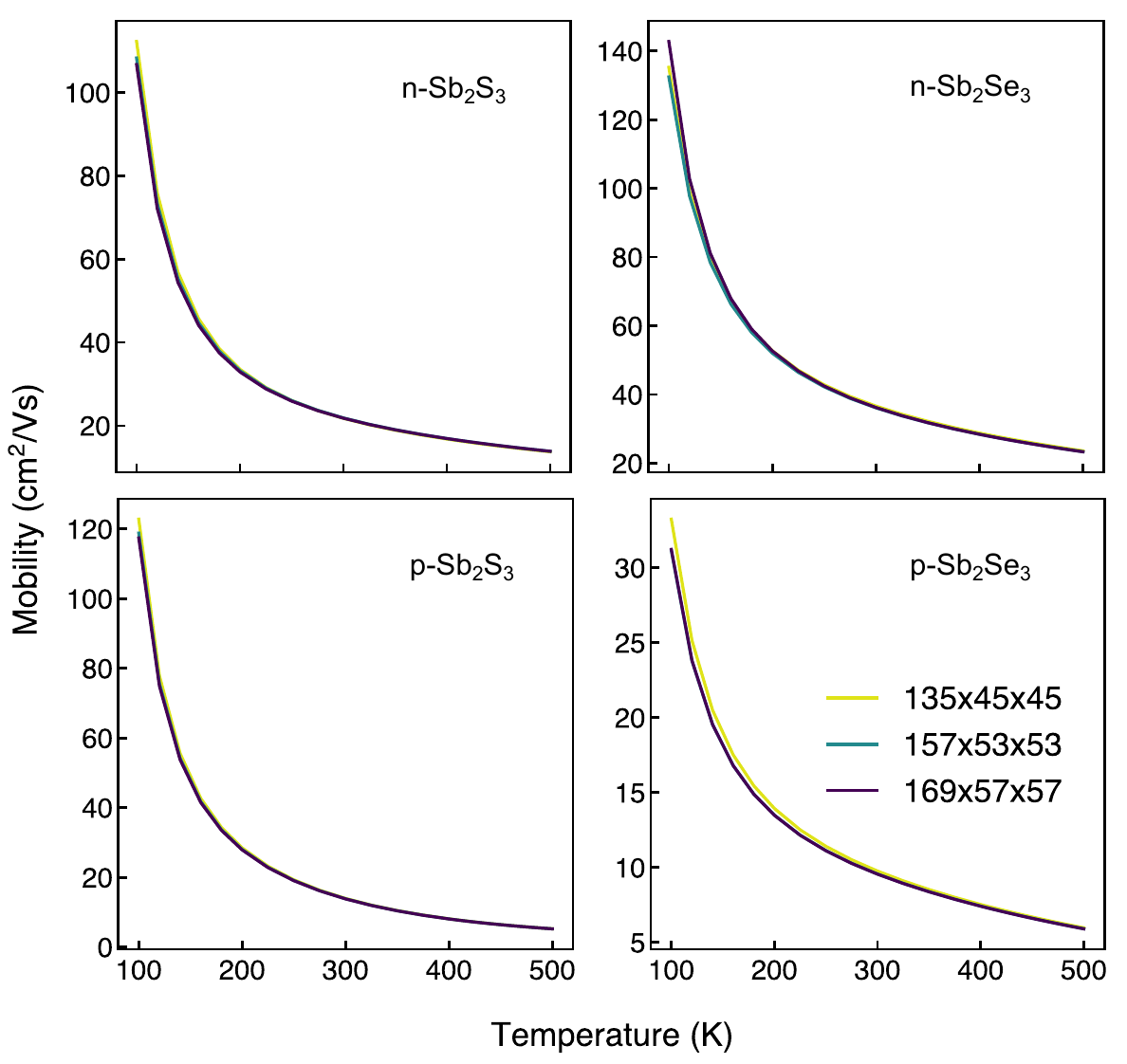}} \\
    \caption{The convergence of mobility in \ce{Sb2X3} under different \textit{k}-point meshes. The defect concentration is set to be \SI{e14}{\conc} and the temperature is set to be \SI{300}{\kelvin}.}
    \label{fig_con}
\end{figure}

The \textit{k}-point meshes used to calculate transport properties were tested (shown in Fig. \ref{fig_con}) and a \textit{k}-point mesh of 169$\times$57$\times$57 is used for all calculations. The carrier concentration was set to \SI{e13}{\conc} according to previous experimental results in \ce{Sb2X3} \cite{chen2017characterization,liu2016green,zhou2014solution,yuan2016rapid,li2021defect,chalapathi2020influence,black1957electrical}. The calculated effective phonon frequency is 3.49 for \ce{Sb2S3} and 2.57 for \ce{Sb2Se3}. The calculated deformation potentials, elastic constants and dielectric constants are shown in Table \ref{tab_def}, \ref{tab_ela} and \ref{tab_die}, respectively.

\begin{table}[ht]
\caption{Calculated deformation potentials (D, eV) for the upper valence and lower conduction bands of \ce{Sb2S3} and \ce{Sb2Se3}}
  \label{tab_def}
  \begin{tabular*}{\textwidth}{@{\extracolsep{\fill}}cccccc}
\hline
Material & \multicolumn{2}{l}{} & D$_{XX}$ & D$_{YY}$ & D$_{ZZ}$ \\ \hline
\multirow{6}{*}{\ce{Sb2S3}} & \multirow{3}{*}{VBM} & D$_{XX}$ & 5.41 & 0.26 & 0.07 \\
 &  & D$_{YY}$ & 0.26 & 0.10 & 0.02 \\
 &  & D$_{ZZ}$ & 0.07 & 0.02 & 1.27 \\ \cline{2-6}
 & \multirow{3}{*}{CBM} & D$_{XX}$ & 5.26 & 0.42 & 0.17 \\
 &  & D$_{YY}$ & 0.42 & 2.43 & 3.35 \\
 &  & D$_{ZZ}$ & 0.17 & 3.35 & 2.62 \\ \hline
\multirow{6}{*}{\ce{Sb2Se3}} & \multirow{3}{*}{VBM} & D$_{XX}$ & 0.53 & 0.16 & 0.05 \\
 &  & D$_{YY}$ & 0.16 & 2.86 & 0.03 \\
 &  & D$_{ZZ}$ & 0.05 & 0.03 & 2.47 \\ \cline{2-6}
 & \multirow{3}{*}{CBM} & D$_{XX}$ & 3.31 & 0.36 & 0.09 \\
 &  & D$_{YY}$ & 0.36 & 0.39 & 0.29 \\
 &  & D$_{ZZ}$ & 0.09 & 0.29 & 1.38 \\ \hline
\end{tabular*}
\end{table}

\begin{table}[ht]
 \caption{Calculated elastic constants (in GPa) of \ce{Sb2S3} and \ce{Sb2Se3}}
  \label{tab_ela}
\begin{tabular*}{\textwidth}{@{\extracolsep{\fill}}c@{\extracolsep{\fill}}c@{\extracolsep{\fill}}c@{\extracolsep{\fill}}c@{\extracolsep{\fill}}c@{\extracolsep{\fill}}c@{\extracolsep{\fill}}c@{\extracolsep{\fill}}c}
\hline
Material &  & C$_{XX}$ & C$_{YY}$ & C$_{ZZ}$ & C$_{XY}$  & C$_{YZ}$ & C$_{ZX}$ \\ \hline
\multirow{6}{*}{\ce{Sb2S3}} & C$_{XX}$ & 93.75 & 28.00 & 18.50 & 0.00 & 0.00 & 0.00 \\
 & C$_{YY}$ & 28.00 & 57.25 & 15.39 & 0.00 & 0.00 & 0.00 \\
 & C$_{ZZ}$ & 18.50 & 15.39 & 37.69 & 0.00 & 0.00 & 0.00 \\
 & C$_{XY}$ & 0.00 & 0.00 & 0.00 & 31.68 & 0.00 & 0.00 \\
 & C$_{YZ}$ & 0.00 & 0.00 & 0.00 & 0.00 & 17.11 & 0.00 \\
 & C$_{ZX}$ & 0.00 & 0.00 & 0.00 & 0.00 & 0.00 & 8.77 \\ \hline
\multirow{6}{*}{\ce{Sb2Se3}} & C$_{XX}$ & 77.15 & 25.63 & 17.11 & 0.00 & 0.00 & 0.00 \\
 & C$_{YY}$ & 25.63 & 54.15 & 17.03 & 0.00 & 0.00 & 0.00 \\
 & C$_{ZZ}$ & 17.11 & 17.03 & 31.75 & 0.00 & 0.00 & 0.00 \\
 & C$_{XY}$ & 0.00 & 0.00 & 0.00 & 23.42 & 0.00 & 0.00 \\
 & C$_{YZ}$ & 0.00 & 0.00 & 0.00 & 0.00 & 18.41 & 0.00 \\
 & C$_{ZX}$ & 0.00 & 0.00 & 0.00 & 0.00 & 0.00 & 5.08 \\ \hline
\end{tabular*}
\end{table}

\begin{table}[ht]
\caption{Calculated static (\textit{\textepsilon}$_0$) and high-frequency (\textit{\textepsilon}$_{\infty}$) dielectric constants of \ce{Sb2S3} and \ce{Sb2Se3}}
 \label{tab_die}
\begin{tabular*}{1\textwidth}{@{\extracolsep{\fill}}c@{\extracolsep{\fill}}c@{\extracolsep{\fill}}c@{\extracolsep{\fill}}c@{\extracolsep{\fill}}c@{\extracolsep{\fill}}c@{\extracolsep{\fill}}c}
 \hline
 \multirow{2}{*}{Material} & \multicolumn{3}{c}{\textit{\textepsilon}$_0$} & \multicolumn{3}{c}{\textit{\textepsilon}$_{\infty}$} \\ 
  & \textit{x}     & \textit{y}      & \textit{z}    & \textit{x}      & \textit{y}     & \textit{z}    \\ \hline
  Sb$_2$S$_3$                               & 98.94 & 94.21  & 13.14 & 11.55      & 10.97     & 8.25    \\ 
  Sb$_2$Se$_3$                              & 85.64 & 128.18 & 15.00  & 15.11  & 14.92 & 10.53 \\ \hline
\end{tabular*}
\end{table}

\clearpage
\bibliography{References}